\documentclass[a4paper,fleqn,usenatbib]{mnras}

\usepackage[T1]{fontenc}
\usepackage{ae,aecompl}

\usepackage{graphicx}	
\usepackage{amsmath}	
\usepackage{amssymb}	

\usepackage{amsfonts}
\usepackage{array}
\usepackage{epsf}
\usepackage{epstopdf}
\usepackage{gensymb}
\usepackage{graphics}
\usepackage{hyperref}
\usepackage{longtable}
\usepackage{lscape}
\usepackage{multirow}
\usepackage{pifont}
\usepackage{rotating}
\usepackage{cancel}

\def\reff@jnl#1{{\rm#1\/}}

\def\aj{\reff@jnl{AJ}} 
\def\araa{\reff@jnl{ARA\&A}} 
\def\apj{\reff@jnl{ApJ}} 
\def\apjl{\reff@jnl{ApJ}} 
\def\apjs{\reff@jnl{ApJS}} 
\def\ao{\reff@jnl{Appl.Optics}} 
\def\apss{\reff@jnl{Ap\&SS}} 
\def\aap{\reff@jnl{A\&A}} 
\def\aapr{\reff@jnl{A\&A~Rev.}} 
\def\aaps{\reff@jnl{A\&AS}} 
\def\azh{\reff@jnl{AZh}} 
\def\baas{\reff@jnl{BAAS}} 
\def\jrasc{\reff@jnl{JRASC}} 
\def\memras{\reff@jnl{MmRAS}} 
\def\mnras{\reff@jnl{MNRAS}} 
\def\pra{\reff@jnl{Phys.Rev.A}} 
\def\prb{\reff@jnl{Phys.Rev.B}} 
\def\prc{\reff@jnl{Phys.Rev.C}} 
\def\prd{\reff@jnl{Phys.Rev.D}} 
\def\prl{\reff@jnl{Phys.Rev.Lett}} 
\def\pasp{\reff@jnl{PASP}} 
\def\pasj{\reff@jnl{PASJ}} 
\def\qjras{\reff@jnl{QJRAS}} 
\def\skytel{\reff@jnl{S\&T}} 
\def\solphys{\reff@jnl{Solar~Phys.}} 
\def\sovast{\reff@jnl{Soviet~Ast.}} 
\def\ssr{\reff@jnl{Space~Sci.Rev.}} 
\def\zap{\reff@jnl{ZAp}} 
\def\nat{\reff@jnl{Nature}} 
\def\icarus{\reff@jnl{Icarus}} 
\def\acta{\reff@jnl{Acta Astron.}} 

\title[TTV and transmission spectroscopy of GJ3470b]{Transit timing variation and transmission spectroscopy analyses of the hot Neptune GJ3470b}

\author[S. Awiphan et al.]{S. Awiphan$^{1}$\thanks{E-mail: supachai@narit.or.th}, E. Kerins$^{1}$\thanks{E-mail: eamonn.kerins@manchester.ac.uk}, S. Pichadee$^{2}$, S. Komonjinda$^{2}$, V. S. Dhillon$^{3,4}$, 
\newauthor W. Rujopakarn$^{5,6}$, S. Poshyachinda$^{7}$, T. R. Marsh$^{8}$, D. E. Reichart$^{9}$,
\newauthor K. M. Ivarsen$^{9}$, and J. B. Haislip$^{9}$\\
$^{1}$Jodrell Bank Centre for Astrophysics, School of Physics and Astronomy, University of Manchester, Oxford Road, Manchester M13 9PL, UK\\
$^{2}$Department of Physics and Material Sciences, Faculty of Science, Chiang Mai University, Chiang Mai 50200, Thailand\\
$^{3}$Department of Physics and Astronomy, University of Sheffield, Sheffield S3 7RH, UK\\
$^{4}$Instituto de Astrofisica de Canarias, La Laguna, Tenerife 38205, Spain\\
$^{5}$Department of Physics, Faculty of Science, Chulalongkorn University, 254 Phayathai Road, Pathumwan, Bangkok 10330, Thailand\\
$^{6}$Kavli Institute for the Physics and Mathematics of the Universe (WPI), The University of Tokyo Institutes for Advanced Study,\\
The University of Tokyo, Kashiwa, Chiba 277-8583, Japan\\
$^{7}$National Astronomical Research Institute of Thailand, 191 Siriphanich Bldg, Huay Kaew Road, Muang District, Chiang Mai 50200, Thailand\\
$^{8}$Department of Physics, University of Warwick, Coventry CV4 7AL, UK\\
$^{9}$Department of Physics and Astronomy, University of North Carolina at Chapel Hill, Chapel Hill, NC 27599, USA}

\date{Accepted XXX. Received YYY; in original form ZZZ}

\pubyear{2016}

\begin{document}
\label{firstpage}
\pagerange{\pageref{firstpage}--\pageref{lastpage}}
\maketitle


\begin{abstract}
GJ3470b is a hot Neptune exoplanet orbiting an M dwarf and the first sub-Jovian planet to exhibit Rayleigh scattering. We present transit timing variation (TTV) and transmission spectroscopy analyses of multi-wavelength optical photometry from 2.4-m and 0.5-m telescopes at the Thai National Observatory, and the 0.6-m PROMPT-8 telescope in Chile. Our TTV analysis allows us to place an upper mass limit for a second planet in the system. The presence of a hot Jupiter with a period of less than 10~days or a planet with an orbital period between 2.5 and 4.0~days are excluded. Combined optical and near-infrared transmission spectroscopy favour an H/He dominated haze (mean molecular weight $1.08\pm0.20$) with high particle abundance at high altitude. We also argue that previous near-infrared data favour the presence of methane in the atmosphere of GJ3470b.
\end{abstract}

\begin{keywords}
techniques: photometric - eclipses - planetary systems - planets and satellites: atmospheres - planets and satellites: individual: GJ 3470b - stars: late-type
\end{keywords}



\section{Introduction}
Over the last decade, the search for and study of exoplanets has been one of the most dynamic research fields of modern astronomy. To date, more than 3,200 planets have been confirmed by various methods\footnote{See \href{http://exoplanet.eu/}{\texttt{http://exoplanet.eu/}}}, including over 2,000 by {\it Kepler} using the transit method \citep{mor2016}. The transit method can detect planets ranging in size from Earth to larger than Jupiter. Additionally, the transit timing variation (TTV) method has been used to find at least 10 additional exoplanets and hundreds of candidates \citep{ago2005,hol2005,hol2010,for2012,for2012b,fab2012,ste2012,ste2012b,ste2013,maz2013}.

In addition to the discovery of new exoplanets, the characterization of planetary interiors and atmospheres is a rapidly developing area. One method that is used to study planetary atmospheres is transmission spectroscopy, which measures the variation of transit depth with wavelength \citep{sea2000}. From the transmission spectroscopy technique, the absorption spectrum and the composition of the planetary atmosphere can be deduced. This method has been applied to several transiting exoplanets, for example, HD189733b \citep{gri2008,swa2010}, GJ1214b \citep{bea2010,kre2014} and GJ436b \citep{knu2014}. 

{\it Kepler} has discovered more than 4,000 planetary candidates \citep{cou2015}. Most of them are super-Earth and Neptune-size candidates (1.25 - 6 $R_{\oplus}$), which confirm the large fraction of small planets within the exoplanet population \citep{howa2012,cou2015}. Exoplanets within this radius range likely comprise solid core (super-Earths), H/He gas and volatile envelopes (Neptune-like exoplanets). 

In order to classify the transition between super-Earths and Neptune-like exoplanets, \cite{lop2014} suggested that planets with a radius larger than about 1.75 $R_{\oplus}$ have H/He envelopes. Nevertheless, the transition point may vary between 1.5 and 2.0 $R_{\oplus}$ \citep{wei2014,mar2014b}. Instead of classification by planetary radius, \cite{raf2011} studied envelope accretion of Neptune-like planets and suggested a mass transition limit at 10 $M_{\oplus}$ or larger for close-in planets. \cite{pis2015} also suggested that the minimum core mass to form a Neptune-like planet is $\sim$8 $M_{\oplus}$ at 5 au and $\sim$5 $M_{\oplus}$ at 100 au. However, in the case of low-density super-Earth sized exoplanets, the mass of the planet alone cannot be used for classification [e.g. Kepler-11f \citep{lis2011} and Kepler-51b \citep{mas2014}]. Planetary average densities are also unable to confirm the transition between super-Earths and Neptune-like exoplanets, due to the broad range of detected planet densities that overlap the transition range \citep{how2014}.

One feature that can be used to determine the solid core to H/He envelope transition regime is the amount of hydrogen and helium in the planet envelope, which can be obtained from transmission spectroscopy measurements \citep{mil2010}. An exoplanet with an H/He-rich atmosphere has a large atmospheric scale height, $h$, due to a low mean molecular weight. 

\begin{equation}
h = \frac{k_{b}T_{p}}{\mu g_{p}} \ ,
\label{Equation:ScaleHeight}
\end{equation}
where $k_{b}$, $T_{p}$, $\mu$ and $g_{p}$ are the Boltzmann constant, the planet equilibrium temperature, the mean molecular weight of the planetary atmosphere and the planet surface gravity, respectively.

In the case of a cloudless atmosphere, absorption lines from volatile molecules at near-infrared wavelengths should be detectable. On the other hand, an exoplanet with low atmospheric H/He abundance provides a smaller scale height and flatter transmission spectrum. However, the presence of high-altitude hazes might cause difficulty in distinguishing between H/He-rich and volatile-rich envelopes, because the haze can hide molecules in the lower atmosphere and produce near-infrared transmission spectra dominated by Mie scattering \citep{how2012}.

An exoplanet around a nearby M dwarf is favourable for transmission spectroscopy, due to its large planet-host radius ratio. In this paper, a transmission spectroscopy study of GJ3470b, a hot Neptune orbiting around an nearby M dwarf, is presented. GJ3470b was first discovered with the HARPS spectrograph and confirmed with follow-up transit observations with the TRAPPIST, Euler and NITES telescopes \citep{bon2012}. GJ3470b is a good target for transmission spectroscopy because it has a large change in transit depth with wavelength due to its large atmospheric opacity \citep{ben2014}. GJ3470b is also the first sub-Jovian planet that shows a significant Rayleigh scattering slope \citep{nas2013}. To date, GJ3470b has been observed at several optical and near-infrared wavelengths \citep{fuk2013,cro2013,dem2013,nas2013,bid2014,ehr2014,dra2015}.

\cite{fuk2013} observed GJ3470b with simultaneous optical and near-infrared observations with the 0.5 m MITSuME and 1.88 m telescopes at Okayama Astrophysical Observatory. They suggested that GJ3470b has a cloud-free atmosphere. \cite{nas2013} combined their optical observations with the Large Binocular Telescope (LBT) \citep{dem2013,fuk2013}. Their result suggests that the GJ3470b atmosphere is cloud-free with a high-altitude haze of tholins. They also found a strong Rayleigh-scattering slope at visible wavelengths.

However, \cite{cro2013} performed an observation with the MOSFIRE spectrograph at the Keck I telescope. They concluded that the GJ3470b atmosphere provides a flat transmission spectrum which indicates methane-poor, metal-rich, optically-thick clouds or a hazy atmosphere. \cite{bid2014} presented 12 new broad-band optical transit observations and concluded that GJ3470b has a hydrogen-rich atmosphere exhibiting a strong Rayleigh-scattering slope from a hazy atmosphere with 50 times solar abundance. 

A recent study by \cite{ehr2014} with the Wide Field Camera-3 (WFC3) on the {\it Hubble} Space Telescope (HST) in the near-infrared also suggested that GJ3470b is dominated by a cloudy hydrogen-rich atmosphere with extremely low water volume mixing ratios ($<$1~ppm). \cite{dra2015} provided shorter-wavelength transmission spectroscopy results with the LCOGT network and the Kuiper telescope in order to verify the Rayleigh scattering signal in the 400-900 nm region. They found a strong Rayleigh scattering slope that indicates an H/He atmosphere with hazes, as in previous studies. 

In this paper, TTV and transmission spectroscopy analyses from additional multi-wavelength optical observations of GJ3470b are presented. These data are combined with previous studies to provide improved measurements of the physical characteristics and ephemeris of GJ3470b. We use these to deduce the first constraints on the characteristics of additional planets in the system. In Section 2, the observational and data analysis details are presented. The mid-transit times from the analysis, together with results from previous studies are used to analyze timing variations and to place an upper mass limit for additional planets in Section 3. In Section 4, the atmosphere of GJ3470b is analyzed using the transmission spectroscopy technique and finally, in Section 5, we present our conclusions.

\section{Observations and data analysis}

\renewcommand{\thefootnote}{\fnsymbol{footnote}}

\begin{table*}
	\centering
	\caption{Observation details for our transmission spectroscopy measurements of GJ3470b.}
	\begin{tabular}{lcccc}
	\hline
	\hline
	\textbf{Observation date} & \textbf{Telescope} & \textbf{Filter} & \textbf{Exposure (s)} & \textbf{Number of images} \\
	\hline
	2013 December 17 & TNO 0.5 m & Cousins-$R$ & 20.0,15.0\footnotemark[1] & 584 \\
	2014 January 06	& TNO 0.5 m & Cousins-$R$ & 15.0 & 500 \\
	2014 January 10	& PROMPT-8 & Cousins-$R$ & 10.0 & 425 \\
	2014 March 04 & TNT & Sloan $z'$ & 5.65 & 1883 \\
	2014 March 14 & TNT & Sloan $r'$ & 8.29 & 1777 \\
	2014 April 03 & TNT & Sloan $z'$ & 5.65 & 1254 \\
	2015 January 22	& PROMPT-8 & Cousins-$R$ & 10.0 & 449 \\
	2015 March 06 & TNT & Sloan $r'$ & 3.13 & 1500 \\
	2015 March 16 & TNT & Sloan $i'$ & 3.13 & 2755 \\
	2016 March 17 & TNT & Sloan $g'$ & 14.85 & 550 \\
	\hline
	\multicolumn{5}{l}{$^{\star}$: The exposure time of the first 296 images is 20 s and the exposure time of the last 288 images is 15 s.} \\
	\end{tabular} \\
	\label{Table:ObservationLog}
\end{table*}

\subsection{Photometric observations}

Photometric observations of exoplanet GJ3470b were conducted between 2013 December and 2016 March. We obtained 10 transits, including 6 full transits and 4 partial transits. The UT date of the mid-transit time, instrument, filter, exposure time and number of frames in each observation are described below and are listed in Table \ref{Table:ObservationLog}. The four-minute binned light curves are shown in Fig \ref{Figure:LightCurve}.

\subsubsection{0.5 m telescope at Thai National Observatory}

Two full transit observations of GJ3470b were obtained through a Cousins-$R$ filter using an Apogee Altra U9000 3056$\times$3056 pixels CCD camera attached to the 0.5 m Schmidt-Cassegrain Telescope located at Thai National Observatory (TNO), Thailand. The field-of-view of each image is 58$\times$58 arcmin$^{2}$. For the first observation on 2013 December 17, the exposure time was set to be 20 s during the first half of observations (296 images) and 15 s during the second half (288 images) due to the variation in seeing at the site. On 2014 January 6, observations with a 20 s exposure time were obtained. The exposure overhead due to readout is $\sim10$ s.

\subsubsection{PROMPT 8 telescope (0.6 m)}

We observed two full transits with PROMPT 8, a 0.6 m robotic telescope at Cerro Tololo Inter-American Observatory (CTIO), Chile, with a 2048 $\times$ 2048 pixel CCD camera with a scale of 0.624 arcsec/pixel. The observations were performed through a Cousins-$R$ filter on 2014 January 10 and 2015 January 22 with 10 s exposures and an $\sim20$ s overhead between each.

\subsubsection{Thai National Telescope (2.4 m)}

We conducted photometric observations of GJ3470 with ULTRASPEC \citep{dhi2014}, a 1k$\times$1k pixels high-speed frame-transfer EMCCD camera, on the 2.4 m Thai National Telescope (TNT) at TNO during the 2013-16 observing seasons. The camera has a field of view 7.68$\times$7.68 arcmin$^{2}$. The readout time between exposures is only 14 ms. Optical multi-wavelength observations though the $z'$, $i'$, $r'$ and $g'$ filters were performed on separate night.

The host star, GJ3470, is an M-dwarf. Therefore, in Fig \ref{Figure:LightCurve}, the $g'$ band light curve provides smaller signal-to-noise ratio compared to other light curves. Although the $g'$ filter light curve shows a large scatter, it is still an important inclusion for planetary atmosphere modelling, especially for Rayleigh scattering curve fitting (Section \ref{Section:Rayleigh}).

\subsection{Light curve analysis}

In the following work, the calibration was carried out using the DAOPHOT package and the photometry was carried out using Python scripts which perform aperture photometry. In order to fit the light curves, we use the Transit Analysis Package ({\texttt{TAP}, \citeauthor{gaz2012} 2012), a set of IDL routines which employs the Markov Chain Monte Carlo (MCMC) technique of \cite{man2002}. 

We combined our 10 light curves with 4 light curves from \cite{bon2012}, in order to fit their mid transit times. Although \cite{bid2014} re-analyzed and fit the mid-transit times of the \cite{bon2012} data, they did not include the 12th April 2012 transit. 

We set scales for semi-major axis ($a/R_{*}$), period ($P$) and inclination ($i$) to be consistent for all light curves. The planet-star radius ratio ($R_{p}/R_{*}$) and quadratic limb darking coefficients are taken to be filter dependent. The mid-transit times ($T_{0}$) observed at the same epoch are also fixed to be the same value. From the previous studies, the eccentricity of the system is less than 0.051 \citep{bon2012}. Therefore, in this work, a circular orbit is assumed. 1,000,000  MCMC steps are performed and the best fits from \texttt{TAP} are shown in Fig \ref{Figure:LightCurve}. 

The orbital elements calculated by \texttt{TAP} are compared with the results from previous studies in Table \ref{Table:Element}. The results from \texttt{TAP} provide a compatible planetary orbital period, inclination and scaled semi-major axis that agree to within 2-$\sigma$ with results from previous studies. Table \ref{Table:LimbDarkening} compares the quadratic limb darkening coefficients ($u_{1}$ and $u_{2}$) with the values from the \cite{cla2011} catalogue, which is based on the PHOENIX model\footnote{See \href{http://phoenix.ens-lyon.fr/simulator/}{\texttt{http://phoenix.ens-lyon.fr/simulator/}}}, with both least-square and flux conservation fitting methods. We use the limb darkening coefficients of a star with stellar temperature $T_{\textup{eff}} = 3500$ K, surface gravity $\log(g_{*}) = 4.5$ and metallicity [Fe/H] = 0.2, which is the nearest grid point ( $T_{\textup{eff}} = 3600\pm100$ K, $\log(g_{*}) = 4.658\pm0.035$ and [Fe/H] = $0.20\pm0.10$ \citep{dem2013}). The \cite{cla2011} catalogue provides compatible (within 2-$\sigma$ variation) limb darkening coefficients with the best-fitting coefficients from \texttt{TAP}.

\begin{figure*}
	\begin{center}
	\includegraphics[width=1.0\textwidth]{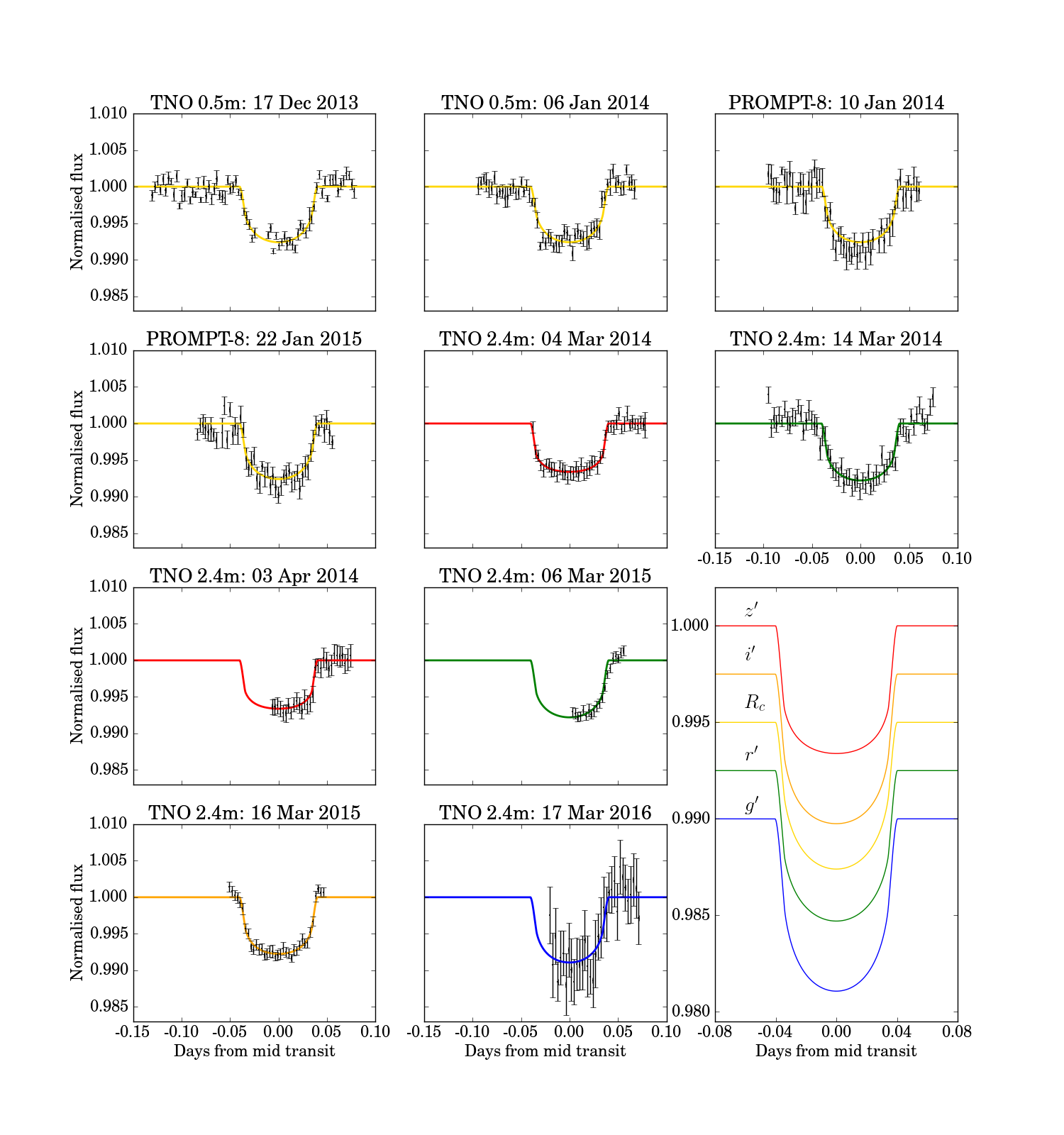}
	\end{center}
	\caption[Light curves of GJ3470b with 4 min binning.]{Light curves of GJ3470b with 4 min binning and with the best-fitting model from the \texttt{TAP} analysis. The best-fitting model light curves in all filters are shown in the bottom right panel with arbitrary off-sets (-0.0025, -0.0050, -0.0075 and -0.0100). Thick yellow, red, orange, green and blue lines represent the best-fitting model in Cousins-$R$, $z'$, $i'$, $r'$ and $g'$ filters, respectively.} 
	\label{Figure:LightCurve}
\end{figure*}
\newpage
\indent 

\begin{landscape}
	\begin{table}
	\centering
	\caption[GJ3470b orbital elements.]{GJ3470b orbital elements from our \texttt{TAP} analysis and from previous studies.}
	\begin{tabular}{lccccccc}
	\hline
	\hline
	\textbf{Reference} & \textbf{Orbital period} & \textbf{Inclination} & \textbf{$a/R_{*}$} & \textbf{$R_{p}/R_{*}$} & \textbf{$u_{1}$} & \textbf{$u_{2}$} & \textbf{Filter}	 \\
	&(Days)&(Degree) \\
	\hline
	\cite{bon2012} & $3.33714 \pm 0.00017$& $>88.8$& $14.9 \pm 1.2$ & $0.0755 \pm 0.0031$ & $0.04$ $^{[1]}$ & $0.19$ $^{[1]}$ & Gunn-$Z$  \\
 & & & & &$0.38$ $^{[1]}$ & $0.40$ $^{[1]}$ & No filter \\          
	\hline
	\cite{dem2013} & $3.33665\pm0.00005$& $88.3^{\mathrm {+0.5}}_{\mathrm {-0.4}}$ & $13.42^{\mathrm {+0.55}}_{\mathrm {-0.53}}$ & $0.07798^{\mathrm {+0.00046}}_{\mathrm {-0.00045}}$ & $0.033\pm0.015$ & $0.181\pm0.10$&IRAC4.5$\mu$m \\
	\hline
	\cite{fuk2013} & $3.336648\pm0.000005$&-&$14.02^{\mathrm {+0.33}}_{\mathrm {-0.39}}$ & $0.07577^{\mathrm {+0.00072}}_{\mathrm {-0.00075}}$ & $0.137^{\mathrm {+0.077}}_{\mathrm {-0.073}}$ & $0.255$ $^{[1]}$ & $J$\\
	& & & &$0.0802\pm0.0013$ & $0.19\pm0.11$ & $0.338$ $^{[1]}$ & $I_c$\\
	& & & &$0.0776\pm0.0038$ & $0.25\pm0.15$ & $0.322$ $^{[1]}$ & $R_c$\\
	& & & &$0.0809\pm0.0031$ & $0.486$ $^{[1]}$ & $0.289$ $^{[1]}$ & $g^{'}$\\
	\hline
	\cite{nas2013} & $3.336649\pm0.000002$& $88.12^{\mathrm {+0.34}}_{\mathrm {-0.30}}$ & - & $0.07484^{\mathrm {+0.00052}}_{\mathrm {-0.00048}}$ & $0.25\pm0.04$ & $0.49\pm0.03$&F972N20 \\
& & & &$0.0821\pm0.0013$ & $0.30\pm0.03$ & $0.29\pm0.03$ & $U$\\
	\hline
	\cite{cro2013} & $3.336665\pm0.000002$ & $88.98^{\mathrm {+0.94}}_{\mathrm {-1.25}}$ & - & $0.0789^{\mathrm {+0.0021}}_{\mathrm {-0.0019}}$ & $-0.351^{\mathrm {+0.025}}_{\mathrm {-0.023}}$ & $-0.889^{\mathrm {+0.051}}_{\mathrm {-0.052}}$&2.09-2.36 $\mu$m \\	
	\hline
	\cite{bid2014} & $3.336648^{\mathrm {+0.0000043}}_{\mathrm {-0.0000033}}$ & $88.88^{\mathrm {+0.44}}_{\mathrm {-0.45}}$ & $13.94^{\mathrm {+0.44}}_{\mathrm {-0.49}}$ & $0.0766^{\mathrm {+0.0019}}_{\mathrm {-0.0020}}$ & $0.017^{\mathrm {+0.014}}_{\mathrm {-0.012}}$ & $0.5030\pm0.0068$&Gunn-$Z$\\ 
	& & & & $0.0766^{\mathrm {+0.0019}}_{\mathrm {-0.0020}}$ & $0.029^{\mathrm {+0.025}}_{\mathrm {-0.018}}$ & $0.5030\pm0.014$&Panstarrs-$Z$\\
	& & & & $0.0765^{\mathrm {+0.0027}}_{\mathrm {-0.0030}}$ & $0.123^{\mathrm {+0.038}}_{\mathrm {-0.047}}$ & $0.488\pm0.020$ & $i^{'}$\\
	& & & & $0.0780^{\mathrm {+0.0015}}_{\mathrm {-0.0016}}$ & $0.070\pm0.025$ & $0.517^{\mathrm {+0.010}}_{\mathrm {-0.0099}}$ & $I$\\
	& & & & $0.0736^{\mathrm {+0.0029}}_{\mathrm {-0.0031}}$ & $0.083^{\mathrm {+0.035}}_{\mathrm {-0.032}}$ & $0.519\pm0.016$&Arizona-$I$\\
	& & & & $0.0803\pm0.0025$ & $0.403^{\mathrm {+0.040}}_{\mathrm {-0.044}}$ & $0.390^{\mathrm {+0.036}}_{\mathrm {-0.038}}$ & $r^{'}$\\	
	& & & & $0.084^{\mathrm {+0.013}}_{\mathrm {-0.016}}$&-&-&Bessel-$B$\\
\hline
	\cite{dra2015} & $3.3366413 \pm 0.0000060$ & - & $12.92^{\mathrm {+0.72}}_{\mathrm {-0.65}}$ & $0.0771^{\mathrm {+0.0012}}_{\mathrm {-0.0011}}$ & $0.123 \pm 0.050$ & $0.489 \pm 0.050$ &$i^{'}$ $^{[2]}$\\
	& & & & $0.0770^{\mathrm {+0.0020}}_{\mathrm {-0.0019}}$ & $0.360 \pm 0.050$ & $0.411 \pm 0.050$ & Harris-$V$\\
	& & & & $0.0833 \pm 0.0019$ & $0.398 \pm 0.050$ & $0.390 \pm 0.050$ & $g^{'}$\\
	& & & & $0.0827^{\mathrm {+0.0022}}_{\mathrm {-0.0020}}$ & $0.421 \pm 0.050$ & $0.398 \pm 0.050$ & Harris-$B$\\
	\hline
This work & $3.3366496^{\mathrm {+0.0000039}}_{\mathrm {-0.0000033}}$ & $89.13^{\mathrm {+0.26}}_{\mathrm {-0.34}}$ & $13.98^{\mathrm {+0.20}}_{\mathrm {-0.28}}$ & $0.0744^{\mathrm {+0.0020}}_{\mathrm {-0.0020}}$ & $0.356^{\mathrm {+0.081}}_{\mathrm {-0.094}}$ & $0.307^{\mathrm {+0.091}}_{\mathrm {-0.112}}$& $z^{'}$\\
	& & & & $0.0785^{\mathrm {+0.0008}}_{\mathrm {-0.0008}}$ & $0.469^{\mathrm {+0.026}}_{\mathrm {-0.046}}$ & $0.350^{\mathrm {+0.031}}_{\mathrm {-0.074}}$& $i^{'}$\\
	& & & & $0.0765^{\mathrm {+0.0017}}_{\mathrm {-0.0015}}$ & $0.585^{\mathrm {+0.023}}_{\mathrm {-0.054}}$ & $0.278^{\mathrm {+0.045}}_{\mathrm {-0.091}}$& $R_c$\\
	& & & & $0.0787^{\mathrm {+0.0016}}_{\mathrm {-0.0022}}$ & $0.540^{\mathrm {+0.079}}_{\mathrm {-0.047}}$ & $0.212^{\mathrm {+0.081}}_{\mathrm {-0.077}}$& $r^{'}$\\
	& & & & $0.0832^{\mathrm {+0.0027}}_{\mathrm {-0.0027}}$ & $0.568^{\mathrm {+0.062}}_{\mathrm {-0.094}}$ & $0.304^{\mathrm {+0.068}}_{\mathrm {-0.099}}$& $g^{'}$\\
	& & & & $0.0752^{\mathrm {+0.0030}}_{\mathrm {-0.0029}}$ & $0.22^{\mathrm {+0.14}}_{\mathrm {-0.11}}$ & $0.15^{\mathrm {+0.17}}_{\mathrm {-0.09}}$& Gunn-$Z$ $^{[3]}$\\
	& & & & $0.0913^{\mathrm {+0.0047}}_{\mathrm -{0.0053}}$ & $0.45^{\mathrm {+0.17}}_{\mathrm {-0.20}}$ & $0.33^{\mathrm {+0.22}}_{\mathrm {-0.20}}$& No filter $^{[3]}$\\
	\hline
	\hline
	\end{tabular} \\
	\label{Table:Element}
	\end{table} 
	\textbf{Remark} \\
	\indent $^{[1]}$: Fixed value. \\
	\indent $^{[2]}$: Re-analyzed \cite{bid2014} data. \\
	\indent $^{[3]}$: Re-analyzed \cite{bon2012} data.
\end{landscape}

\begin{table}
	\setlength{\tabcolsep}{3pt}
	\centering
	\caption[GJ3470 quadratic limb darkening coefficients]{GJ3470 quadratic limb darkening coefficients from our \texttt{TAP} analysis together with predicted coefficients from the models of Claret et al. (2011), with both least-square (L) and flux conservation (F) fitting methods (see the text).}
	\begin{tabular}{lcccccc}
	\hline
	\hline
	\multirow{2}{*}{\textbf{Filter}} & \multicolumn{2}{c}{\textbf{Best fit}} & \multicolumn{2}{c}{\textbf{Claret L}} & \multicolumn{2}{c}{\textbf{Claret F}} \\
	 & $u_{1}$ & $u_{2}$ & $u_{1}$ & $u_{2}$ & $u_{1}$ & $u_{2}$ \\
	\hline
	R & $0.585^{\mathrm {+0.023}}_{\mathrm {-0.054}}$ & $0.278^{\mathrm {+0.045}}_{\mathrm {-0.091}}$ & 0.4998 & 0.2329 & 0.5179 & 0.2101 \\
	$g'$ & $0.568^{\mathrm {+0.062}}_{\mathrm {-0.094}}$ & $0.304^{\mathrm {+0.068}}_{\mathrm {-0.099}}$ & 0.5154 & 0.3046 & 0.5405 & 0.2724 \\
	$r'$ & $0.540^{\mathrm {+0.079}}_{\mathrm {-0.047}}$ & $0.212^{\mathrm {+0.081}}_{\mathrm {-0.077}}$ & 0.5419 & 0.2221 & 0.5572 & 0.2028 \\
	$i'$ & $0.469^{\mathrm {+0.026}}_{\mathrm {-0.046}}$ & $0.350^{\mathrm {+0.031}}_{\mathrm {-0.074}}$ & 0.3782 & 0.2830 & 0.4053 & 0.2486 \\
	$z'$ & $0.356^{\mathrm {+0.081}}_{\mathrm {-0.094}}$ & $0.307^{\mathrm {+0.091}}_{\mathrm {-0.112}}$ & 0.3804 & 0.2361 & 0.2746 & 0.3311 \\
	\hline
	\hline
	\end{tabular}
	\label{Table:LimbDarkening}
\end{table}

\begin{table}
	\centering
	\caption[Mean stellar density of GJ3470.]{Mean stellar density of GJ3470 from our \texttt{TAP} analysis and previous studies.}
 	\begin{tabular}{lc}
	\hline
	\hline
	\textbf{Reference} & \textbf{Stellar density ($\rho_\odot$)} \\
	\hline
	\cite{bon2012}&4.26$\pm$0.53\\
	\cite{dem2013}&2.91$\pm$0.37\\
	\cite{pin2013}&4.25$\pm$0.40\\
	\cite{fuk2013}&3.32$\pm$0.27\\
	\cite{nas2013}&2.74$\pm$0.19\\
	\cite{cro2013}&3.49$\pm$1.13\\
	\cite{bid2014}&3.39$^{\mathrm {+0.30}}_{\mathrm {-0.32}}$\\
	\hline
	This work& 3.30$\pm$0.17\\
	\hline
	\hline
	\end{tabular}
	\label{Table:Density}
\end{table}

\begin{table*}
\centering
\caption{Summary of GJ3470b properties.}
 	\begin{tabular}{lccc}
 	\hline
 	\textbf{Parameter} & \textbf{Symbol} & \textbf{Value} & \textbf{Unit} \\
 	\hline
	\multicolumn{4}{l}{{\it \textbf{Stellar parameters}}} \\
 	Stellar mass & $M_{*}$ & $0.539^{+0.047}_{-0.043}$\footnotemark[1] & $M_\odot$ \\
 	Stellar radius & $R_{*}$ & $0.547\pm0.018$ & $R_\odot$ \\
 	Stellar density & $\rho_{*}$ & $3.30\pm0.17$ & g cm$^{-3}$ \\
 	Stellar surface gravity & $\log (g_{*})$ & $4.695\pm0.046$ & cgs \\
 	Stellar effective temperature & $T_{\textup{eff}}$ & $3600\pm100$\footnotemark[1] & K \\
 	Stellar metallicity & [Fe/H] & $0.20\pm0.10$\footnotemark[1] & \\
	\multicolumn{4}{l}{{\it \textbf{Planetary parameters}}} \\
 	Orbital period & $P$ & $3.3366496^{\mathrm {+0.0000039}}_{\mathrm {-0.0000033}}$ & d \\
 	Orbital inclination & $i$ & $89.13^{\mathrm {+0.26}}_{\mathrm {-0.34}}$ & $\deg$ \\
 	Semi-major axis & $a$ & $0.0355\pm0.0019$ & au \\
 	Epoch of mid-transit (BJD) & $T_{0}$ & 2455983.70421 & d \\
 	Radial velocity amplitude parameter & $K^{'}$ & $13.4\pm1.2$\footnotemark[1] & m s$^{-1}$d$^{1/3}$ \\
 	Planetary mass & $M_{p}$ & $13.9\pm1.5$& $M_\oplus$ \\
 	Planetary radius & $R_{p}$ & $4.57\pm0.18$ & $R_\oplus$ \\
 	Planetary density & $\rho_{p}$ & $0.80\pm0.13$ & g cm$^{-3}$ \\
 	Planetary equilibrium temperature & $T_{p}$ & 497 - 690 & K\\
 	Planetary surface gravity & $\log (g_{p})$ & $2.815\pm0.057$ & cgs \\
 	Planetary atmospheric scale height & $h$ & $760\pm140$ & km \\
 	Planetary atmospheric mean molecular weight & $\mu$ & $1.08\pm0.20$ & \\
 	\hline
 	\hline
 	\multicolumn{4}{l}{$^{\star}$: Adopted value from \cite{dem2013}.} \\
 	\end{tabular}
	\label{Table:Parameter}
\end{table*}

\subsection{Stellar and planetary characterizations}

In order to obtain planetary physical parameters, the parameters of the host star must also be considered. The mean stellar density is calculated by Kepler's third law neglecting planetary mass. From the \texttt{TAP} result, the mean density of GJ3470 is $\rho_{*}=3.30\pm0.17 \rho_{\odot}$. This result is consistent with the value derived by other works (see Table \ref{Table:Density}). 

To find other stellar and planetary parameters, we adopt a stellar mass, $M_{*}=0.539^{+0.047}_{-0.043} M_\odot$ from \cite{dem2013}, which were obtained from the average of the $J-$, $H-$, and $K-$band mass-luminosity (M-L) relations of \cite{del2000}. We also adopt a radial velocity amplitude parameter $K^{'} = 13.4\pm1.2$ m s$^{-1}$d$^{1/3}$ from \cite{dem2013}, where

\begin{equation}
K^{'} = KP^{1/3} = \frac{(2\pi G)^{1/3} M_{p} \textup{sin} i}{(M_{*} + M_{p})^{2/3}}
\label{Equation:RadialVelocity}
\end{equation}
for a circular orbit. In the above equation, $K$ is the radial velocity semi-amplitude, $M_{p}$ is the planet mass, $M_{*}$ is the host mass and $G$ is the gravitational constant.

Combining with the mean density, the calculated radius of the GJ3470 host star is $R_{*}=0.547\pm0.018 R_{\odot}$. The planetary radius is calculated from the planet-star radius ratio, which is wavelength dependent. We use the ratio in the Cousins-$R$ waveband to calculate a radius, $R_{p}=4.57\pm0.18 R_{\oplus}$.

The calculated planetary mass and density are $M_{p} = 13.9\pm1.5 M_{\oplus}$ and $\rho_{p} = 0.80\pm0.13$ g cm$^{-3}$. The range of planetary equilibrium temperature, $T_{p}$, can be derived from the relation,

\begin{equation}
T_{p} = T_{\textup{eff}} \left( \frac{1-A}{4F} \right)^{1/4} \left( \frac{R_{*}}{2a}\right) ^{1/2} 
\label{Equation:Temperature}
\end{equation}
where $T_{\textup{eff}}$ = $3600\pm100$ K is the effective temperature of the host star, $A$ is the Bond albedo (0-0.4) and $F$ is the heat redistribution factor (0.25-0.50) \citep{dem2013,bid2014}. In the calculation, we use $T_{\textup{eff}}$ = $3600$ K and $a/R_{*} = 13.98$ and their uncertainties are not taken into account. The temperature range in our work, due to the possible range of Bond albedo and the heat redistribution factor, is $T_{p}$=497-690 K. The list of all parameters from the analysis is shown in Table~\ref{Table:Parameter}.

\section{Transit timing variations}

\subsection{O-C diagram}
\label{Section:OC}

The measured mid-transit time of a photometric light curve always has some variation due to noise. However, variations can be caused by the gravitational interaction of other objects in the system, such as other exoplanets or exomoons. Therefore, we can use the mid-transit times of GJ3470b to place limits on transit timing variations (TTVs). From the \texttt{TAP} analysis, the mid-transit times of 14 light curves are shown in Table \ref{Table:MidTransit}. 

We use these mid-transit times, and those from previous studies, to plot the epoch of each transit against the observed minus the calculated time ($O-C$) in order to find the TTV of GJ3470b. We perform a linear fit to the $O-C$ diagram to correct GJ3470b's ephemeris. The best linear fit (reduced chi-squared, $\chi^{2}_{r,\textup{L}} = 2.11$) gives a corrected ephemeris of 

\begin{equation}
	T_{0}(E) = 2455983.70421 + 3.33665E \hspace{15pt}(\textup{BJD}) ,
\end{equation}
where $E$ is the number of epochs from the 2012 February 26 transit, the first transit of \cite{bon2012}. In Fig \ref{Figure:OCDiagram}, the $O-C$ diagram shows that there is no significant variation of the mid-transit time. Almost all of them are consistent within 2-$\sigma$.

\begin{table}
	\centering
	\caption[Mid transit of GJ3470b.]{Mid transit time and $O-C$ residuals for GJ3470b from our \texttt{TAP} analysis.}
	\begin{tabular}{lcc}
	\hline
	\hline
	\multirow{2}{*}{\textbf{Observing date}} & \textbf{Mid-transit time (BJD)} & \textbf{$O-C$}\\
	& (BJD-2450000) & (d) \\
	\hline
	26 February 2012\footnotemark[1] & 5983.7015$^{\mathrm {+0.0015}}_{\mathrm {-0.0015}}$ & -0.0026 \\
	07 March 2012\footnotemark[1] & 5993.7152$^{\mathrm {+0.0014}}_{\mathrm {-0.0014}}$ & 0.0012 \\
	12 April 2012\footnotemark[1] & 6030.4177$^{\mathrm {+0.0010}}_{\mathrm {-0.0010}}$ & 0.0006 \\
	17 December 2013 & 6644.3602$^{\mathrm {+0.0011}}_{\mathrm {-0.0014}}$ & -0.0006\\
    06 January 2014 & 6664.3812$^{\mathrm {+0.0017}}_{\mathrm {-0.0017}}$ & 0.0005\\
    10 January 2014 & 6667.7169$^{\mathrm {+0.0021}}_{\mathrm {-0.0021}}$ & -0.0004 \\
    04 March 2014 & 6721.1038$^{\mathrm {+0.0004}}_{\mathrm {-0.0005}}$ & 0.0001 \\
    14 March 2014 & 6731.1162$^{\mathrm {+0.0011}}_{\mathrm {-0.0011}}$ & 0.0026 \\
    03 April 2014 & 6751.1332$^{\mathrm {+0.0007}}_{\mathrm {-0.0008}}$ & -0.0004 \\
    22 January 2015 & 7044.7603$^{\mathrm {+0.0019}}_{\mathrm {-0.0024}}$ & 0.0015 \\
    06 March 2015 & 7088.1370$^{\mathrm {+0.0008}}_{\mathrm {-0.0007}}$ & 0.0018 \\
    16 March 2015 & 7098.1455$^{\mathrm {+0.0004}}_{\mathrm {-0.0005}}$ & 0.0003 \\
    17 March 2016 & 7465.1750$^{\mathrm {+0.0017}}_{\mathrm {-0.0014}}$ & -0.0017 \\
	\hline
	\hline
	\multicolumn{2}{l}{$^{\star}$ : Re-analyzed \cite{bon2012} data} \\
	\end{tabular}
	\label{Table:MidTransit}
\end{table}

\begin{figure}
	\begin{center}
	\includegraphics[width=0.5\textwidth]{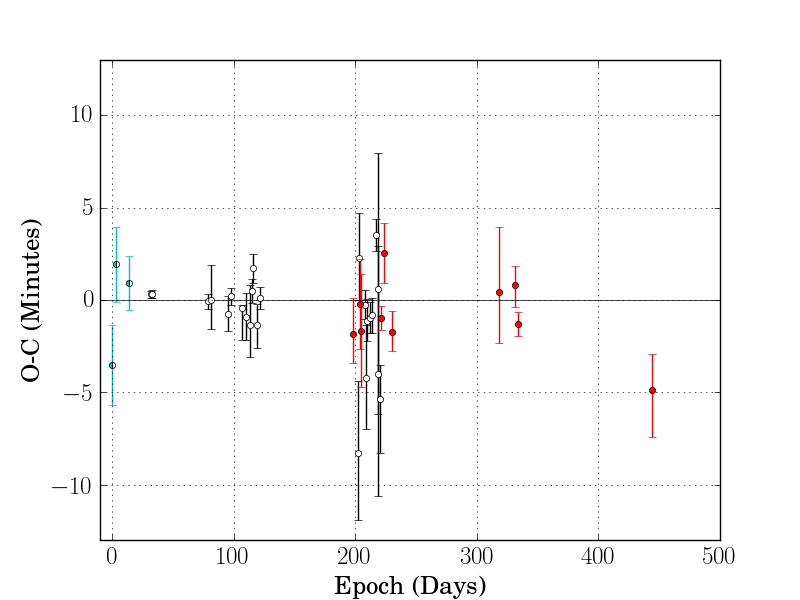}
	\caption{O-C diagram of exoplanet GJ3470b. Epoch = 0 is the transit on 2012 February 26. The red filled, cyan half-filled and unfilled markers represent the mid-transit time from our observations, re-analysed Bonfils et al. (2012) observations and other previous published observations, respectively.}
	\label{Figure:OCDiagram}
	\end{center}
\end{figure}

\subsection{Upper mass limit of the second planet.}

From Section \ref{Section:OC}, the $O-C$ diagram shows no significant TTV signal, indicating that there are no nearby massive objects, which have strong gravitational interaction with GJ3470b. We use this to compute an upper mass limit on a second planet in the system. We assume that the second planet is in a circular orbit that is also coplanar with GJ3470b's orbit. We use the \texttt{TTVFaster} code \citep{ago2016}, which computes the TTV signal from analytic formulae.

We employ two methods to measure the upper mass limit. First, we calculate the TTV signal for the second planet over a mass range from $10^{-1} M_{\oplus}$ to $10^{3} M_{\oplus}$ with $10^{0.01} M_{\oplus}$ steps on a logarithmic scale. We sample a period ratio of the perturbing planet and GJ3470b over a ratio range from 0.30 to 4.50 with 0.01 steps. At each grid point, the initial phase of the perturber is varied between 0 and 2$\pi$ with $\pi/18,000$ steps in order to cover all alignments of the second planet at $E=0$. For each period of the perturber planet, the minimum mass which produces a TTV signal higher than the measured TTV limit is taken to be the upper mass limit for that period. As the highest TTV signal from the $O-C$ diagram is 498 s, upper mass limits corresponding to TTV amplitudes of 400, 500 and 600 s are calculated and shown in Fig \ref{Figure:MinMass}.

The second method uses the reduced chi-squared of the best-fit between the observed TTV signal and the signal from \texttt{TTVFaster}. The grid points and the initial phase of second planet are varied as in the first method. From Section \ref{Section:OC}, the best linear fit using a single-planet model is $\chi^{2}_{r,\textup{L}}=2.11$. We assess the improvement to the fit of introducing a second planet though the delta reduced chi-squared statistic, $\Delta\chi^{2}_{r} = \chi^{2}_{r}-\chi^{2}_{r,\textup{L}}$, where $\chi_{r}^{2}$ is the best fitting TTV model at the given mass and period. In Fig \ref{Figure:MinMass}, $\Delta\chi^{2}_{r}$ is shown as a function of perturber mass and period. The preferred planet models are shown in Fig \ref{Figure:MinMass} as negative valued $\Delta\chi^{2}_{r}$ regions. The best-fitting TTV models are shown with the black dotted line in Fig \ref{Figure:MinMass}, which is produced by averaging over period ratio bins of width 0.05.

Unstable orbit regions are calculated from the mutual Hill sphere between GJ3470b and the perturber. For two-planet systems in coplanar and circular orbits, the boundary of the stable orbit is when the separation of the planets' semi-major axes ($a_{\textup{out}}-a_{\textup{in}}$) is larger than $2\sqrt{3}$ of the mutual Hill sphere \citep{fab2012}

\begin{equation}
r_{H} = \frac{a_{\textup{in}}+a_{\textup{out}}}{2}\left ( \frac{M_{\textup{in}}+M_{\textup{out}}}{3M_{*}} \right )^{1/3} \ .
\label{Equation:Hill}
\end{equation}

In Equation \ref{Equation:Hill}, $a_{\textup{in}}$ and $a_{\textup{out}}$ are the semi-major axis of the inner and outer planets, respectively. The area of unstable orbits is shown by the black shaded region of Fig \ref{Figure:MinMass}. Orbital resonances between GJ3470b and the second planet are shown as vertical lines. In the cases where GJ3470b and the perturber are in a first-order mean motion resonance, the upper mass limits are lower.

From Fig \ref{Figure:MinMass}, the preferred-TTV area with a $\Delta\chi^{2}_{r}$ between -0.7 and -0.4 is shown near the upper mass-limit of 400 s TTV amplitude. A nearby second planet with period between 2.5 and 4.0 days is ruled out by both upper mass limit tests and the mutual Hill sphere area. A Jupiter-mass planet with period less than 10 days is also excluded. From this result, we can conclude that there is no nearby massive planet to GJ3470b.

\begin{figure}
	\begin{center}
	\includegraphics[width=0.55\textwidth]{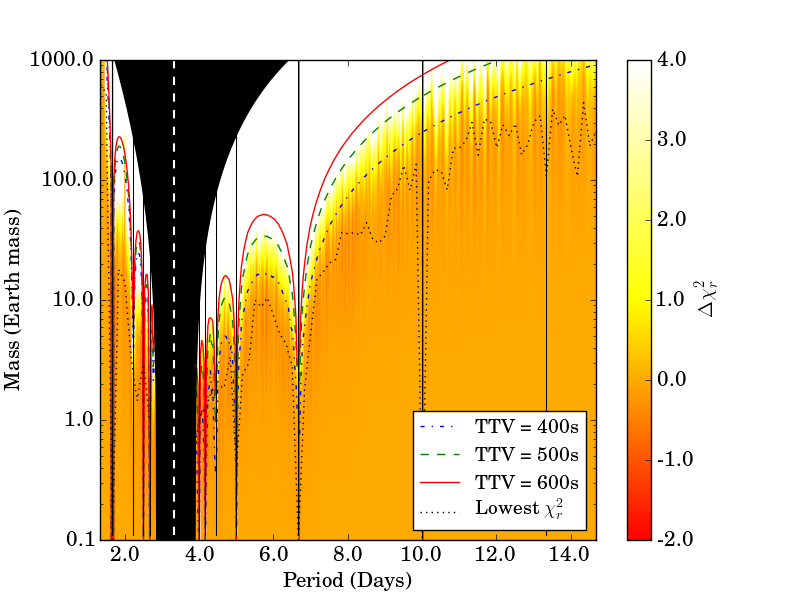}
	\caption[Upper mass limit of a second planet in GJ3470 system.]{Upper mass limit of a second planet in the GJ3470 system. The blue dash-dotted, green dashed and red solid lines represent the upper mass limit for 400, 500 and 600 s TTV amplitudes. The contours show the $\Delta\chi^{2}_{r}$ between the best TTV fit and the best linear fit. The black dotted line presents the best $\Delta\chi^{2}_{r}$ within a 0.05 period ratio bin. From left to right, the black vertical lines show 3:1, 2:1, 3:2, 4:3, 5:6, 4:5, 3:4, 2:3, 1:2, 1:3 and 1:4 resonance periods. The white vertical dashed line shows the orbital period of GJ3470b.}
	\label{Figure:MinMass}
	\end{center}
\end{figure}

\section{Transmission spectroscopy}

\subsection{Rayleigh scattering}
\label{Section:Rayleigh}

Transmission spectroscopy of an exoplanet can be seen as a change in planet-star radius ratio as a function of wavelength. At a wavelength with higher atmospheric absorption, the planet appears larger, due to the opaque atmosphere. For an ideal gas atmosphere in hydrostatic equilibrium, the slope of the planetary radius as a function of wavelength can be expressed as $\textup{d}R_{p}/\textup{d}\ln \lambda = \alpha H$, where $\alpha$ is a scale factor \citep{lec2008}. From the atmospheric scale height relation (Equation \ref{Equation:ScaleHeight}), the mean molecular weight of the planetary atmosphere can be estimated as
\begin{equation}
\mu = \alpha T_{p} k_{B} \left( g_{p}\frac{\textup{d}R_{p}}{\textup{d} \ln \lambda} \right)^{-1} .\
\end{equation}
We adopt a Bond Albedo, $A=0.3$ and an equilibrium temperature, $T = 624\pm25$ K from \cite{dem2013}. We assume the main physical process involved in GJ3470b's atmosphere is Rayleigh scattering ($\alpha = -4$) without atomic or molecular absorption. In order to find $\textup{d}R_{p}/\textup{d}\ln \lambda$, we combine our planet-star radius ratio with previous optical observations. The plot of planet-star radius ratio versus wavelength and the best-fitting model of the GJ3470b data with mean-molecular weight 1.00, 1.50, 2.22 (Jupiter) and 2.61 (Neptune) atmospheres are shown in Fig \ref{Figure:Rayleigh}.

\begin{figure}
	\begin{center}
	\includegraphics[width=0.5\textwidth]{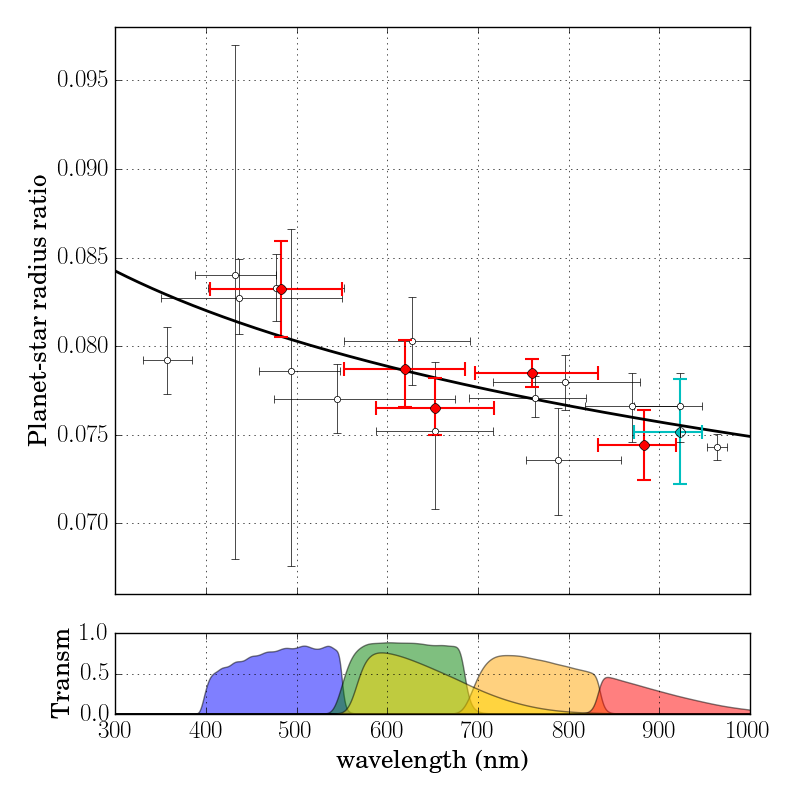}
	\caption[Relation between planet-star radius ratio and wavelength in optical filters.]{(Top) Relation between planet-star radius ratio and wavelength in optical filters with $\textup{d}R_{p}/\textup{d}\ln \lambda$ slope fitting (Black thick line). The blue solid, dashed, dash-dotted and dotted lines represent the best fit with mean-molecular weight 1.00, 1.50, 2.22 (Jupiter) and 2.61 (Neptune), respectively. The markers have the same description as Fig \ref{Figure:OCDiagram}. (Bottom) The bandpass of our filters: $g'$, $r'$, Cousins-$R$, $i'$ and $z'$ band (from left to right). The bandpass colours have the same description as Fig \ref{Figure:LightCurve}.}
	\label{Figure:Rayleigh}
	\end{center}
\end{figure}

From the curve fit to the data in Fig \ref{Figure:Rayleigh}, a low mean molecular weight of $1.08\pm0.20$ is obtained. This low mean molecular weight is consistent with an H/He-dominated atmosphere as in previous studies (\cite{nas2013}: $1.32^{\mathrm {+0.27}}_{\mathrm {-0.19}}$ and \cite{dra2015}: $1.35\pm0.44$).

\subsection{Atmospheric composition}

To determine the atmospheric composition, ideally, a detailed atmospheric model of GJ3470b is required. We rescale the planetary atmosphere models of \cite{how2012}. We use 114 models spanning a range of atmospheric compositions including: sub-solar, solar and super-solar metallicity compositions, as well as single component atmospheres of methane, carbon dioxide and water. For each of these atmospheric models, we test several different assumptions of atmospheric contaminant including: cloud-free and cloudy atmospheres, and polyaceylene and tholin hazes. 

We adopt only models with a planetary equilibrium temperature and mass close to GJ3470b ($T_{p}$ = 700 K, $M_{p} = 10 M_{\oplus}$). The optical planet-star radius ratio data from Section \ref{Section:Rayleigh} and infrared data from \cite{dem2013} and \cite{ehr2014} are used to compute a reduced chi-squared ($\chi^{2}_{r,\textup{atm}}$) between the rescaled models and the data. In Table \ref{Table:AtmosphereModel}, examples of the atmosphere models with their best-fitting $\chi^{2}_{r,\textup{atm}}$ are shown. 

\begin{figure}
	\begin{center}
	\includegraphics[width=0.48\textwidth]{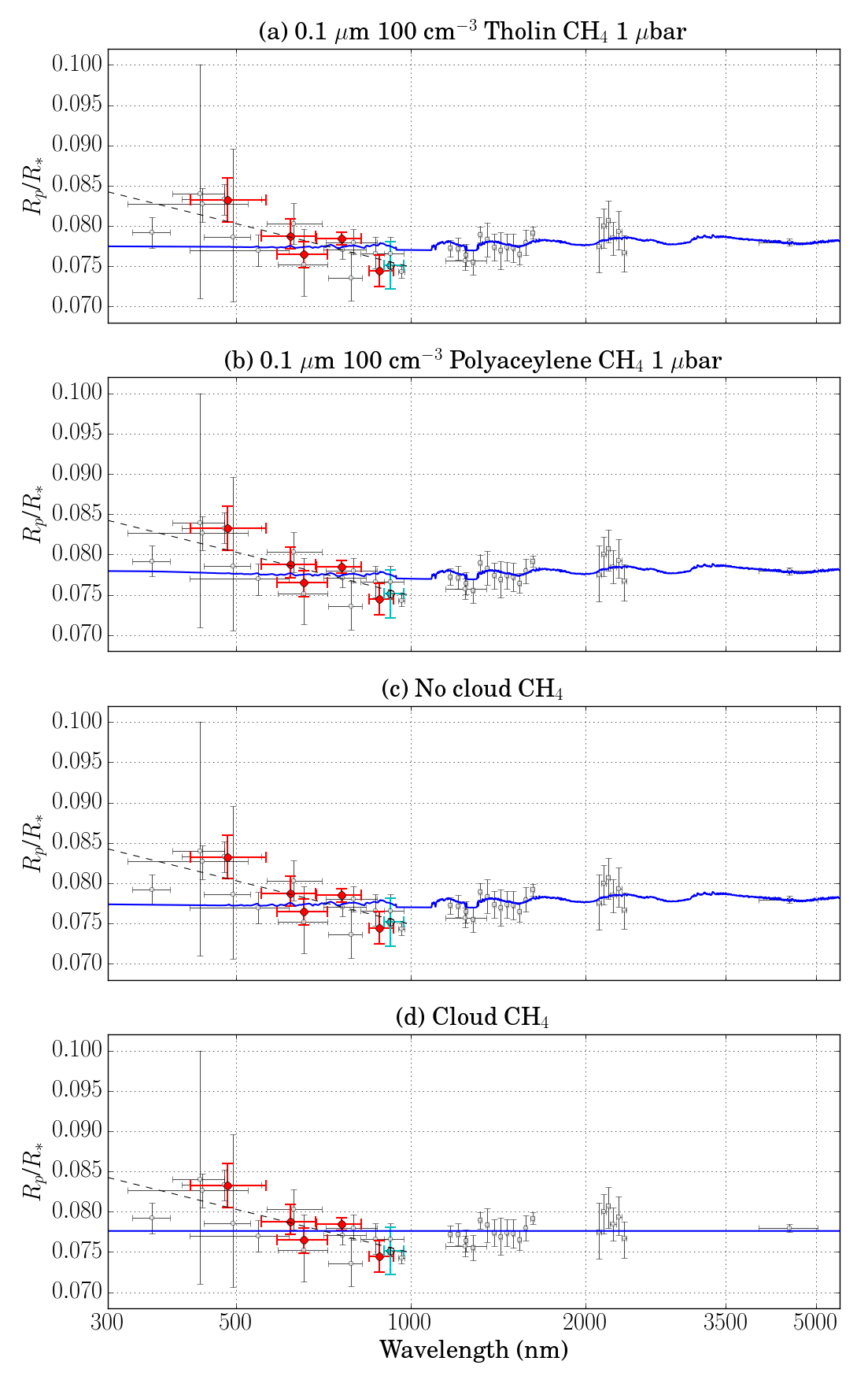}
	\caption[GJ3470b atmosphere models.]{GJ3470b atmosphere models: (a) 0.1 $\mu$m 100 cm$^{-3}$ Tholin with 1-1000 $\mu$bar CH$_{4}$ ($\chi^{2}_{r,\textup{atm}}=1.38$), (b) 0.1 $\mu$m 100 cm$^{-3}$ Polyaceylene with 1-1000 $\mu$bar CH$_{4}$ ($\chi^{2}_{r,\textup{atm}}=1.40$), (c) No cloud CH$_{4}$ ($\chi^{2}_{r,\textup{atm}}=1.49$) and (d) Cloud CH$_{4}$ 1 mbar  ($\chi^{2}_{r,\textup{atm}}=1.64$), with their best $\chi^{2}_{r,\textup{atm}}$ fits (Blue solid lines). The thin dashed lines show the Rayleigh scattering slope fitting at optical wavelength (300-1000 nm, see Section \ref{Section:Rayleigh} for detail). The markers have the same description as Fig \ref{Figure:OCDiagram}.}
	\label{Figure:AtmosphereModel}
	\end{center}
\end{figure}

From the fitting, a CH${_4}$ atmosphere with a 100 cm$^{-3}$ particle abundance haze at 1-1000 $\mu$bar altitude provides the best fit with $\chi^{2}_{r,\textup{atm}}$ = 1.40 or 1.38 for polyacetylene or tholin haze, respectively. At near-infrared wavelengths, CH$_{4}$ models provide the best fit and the data are not compatible with a cloudy atmosphere model. However, the models of \cite{how2012} do not provide an atmosphere with mixed composition (the CH$_{4}$ atmosphere is a 100\% methane atmosphere), which might be the cause of the poor fit to the data at optical wavelengths. From the Rayleigh scattering slope, the mean molecular weight is too low to be a methane dominated atmosphere. Therefore, the H/He dominated haze with high particle abundance, such as high altitude polyacetylene and tholin with a methane contaminant, is preferred. A model atmosphere with a mixed-ratio composition should provide a better description of the GJ3470b atmosphere.

\begin{table*}
	\centering
	\caption[Atmospheric models with the best GJ3470b's atmosphere $\chi^{2}_{r}$ fitting.]{Atmospheric models of Howe \& Burrows (2012) with the best $\chi^{2}_{r,\textup{atm}}$ fit. The highlight best-fitting atmospheric models are shown in Fig \ref{Figure:AtmosphereModel}.}
 	\begin{tabular}{llcc}
	\hline
	\hline
	\textbf{Particle} & \textbf{Composition} & \textbf{Cloud top pressure} & \textbf{$\chi^{2}_{r,\textup{atm}}$} \\
	\hline
	No Cloud & 0.3$\times$ solar & 1 bar & 3.24 \\
	No Cloud & 1$\times$ solar & 1 bar & 4.05 \\
	No Cloud & 3$\times$ solar & 1 bar & 4.64 \\
	\textbf{No Cloud} & \textbf{CH$_{4}$} & \textbf{1 bar} & \textbf{1.49} \\
	No Cloud & CO$_{2}$ & 1 bar & 1.79 \\
	No Cloud & H$_{2}$O & 1 bar & 1.86 \\
	Cloud & 0.3$\times$ solar & 1 mbar & 1.80 \\
	Cloud & 1$\times$ solar & 1 mbar & 2.04 \\
	Cloud & 3$\times$ solar & 1 mbar & 2.33 \\
	\textbf{Cloud} & \textbf{CH$_{4}$} & \textbf{1 mbar} & \textbf{1.64} \\
	Cloud & CO$_{2}$ & 1 mbar & 1.76 \\
	Cloud & H$_{2}$O & 1 mbar & 1.77 \\
	Cloud & 0.3$\times$ solar & 1 $\mu$bar & 1.71 \\
	Cloud & 1$\times$ solar & 1 $\mu$bar & 1.71 \\
	Cloud & 3$\times$ solar & 1 $\mu$bar & 1.71 \\
	Cloud & CH$_{4}$ & 1 $\mu$bar & 1.71 \\
	Cloud & CO$_{2}$ & 1 $\mu$bar & 1.72 \\
	Cloud & H$_{2}$O & 1 $\mu$bar & 1.71 \\
	0.1 $\mu$m 100 cm$^{-3}$ polyaceylene & 0.3$\times$ solar & 1 $\mu$bar & 2.14 \\
	0.1 $\mu$m 100 cm$^{-3}$ polyaceylene & 1$\times$ solar & 1 $\mu$bar & 2.36 \\
	0.1 $\mu$m 100 cm$^{-3}$ polyaceylene & 3$\times$ solar & 1 $\mu$bar & 2.51 \\
	\textbf{0.1 $\mu$m 100 cm$^{-3}$ polyaceylene} & \textbf{CH$_{4}$} & \textbf{1 $\mu$bar} & \textbf{1.40} \\
	0.1 $\mu$m 100 cm$^{-3}$ polyaceylene & CH$_{4}$ & 1 mbar & 1.47 \\
	0.1 $\mu$m 100 cm$^{-3}$ polyaceylene & CO$_{2}$ & 1 $\mu$bar & 1.78 \\
	0.1 $\mu$m 100 cm$^{-3}$ polyaceylene & H$_{2}$O & 1 $\mu$bar & 1.76 \\
	0.1 $\mu$m 1000 cm$^{-3}$ polyaceylene & CH$_{4}$ & 1 $\mu$bar & 1.49 \\
	1.0 $\mu$m 0.01 cm$^{-3}$ polyaceylene & CH$_{4}$ & 1 $\mu$bar & 1.50 \\
	1.0 $\mu$m 0.1 cm$^{-3}$ polyaceylene & CH$_{4}$ & 1 $\mu$bar & 1.47 \\
	0.1 $\mu$m 100 cm$^{-3}$ tholin & 0.3$\times$ solar & 1 $\mu$bar & 2.13 \\
	0.1 $\mu$m 100 cm$^{-3}$ tholin & 1$\times$ solar & 1 $\mu$bar & 2.31 \\
	0.1 $\mu$m 100 cm$^{-3}$ tholin & 3$\times$ solar & 1 $\mu$bar & 2.44 \\
	\textbf{0.1 $\mu$m 100 cm$^{-3}$ tholin} & \textbf{CH$_{4}$} & \textbf{1 $\mu$bar} & \textbf{1.38} \\
	0.1 $\mu$m 100 cm$^{-3}$ tholin & CH$_{4}$ & 1 mbar & 1.46 \\
	0.1 $\mu$m 100 cm$^{-3}$ tholin & CO$_{2}$ & 1 $\mu$bar & 1.78 \\
	0.1 $\mu$m 100 cm$^{-3}$ tholin & H$_{2}$O & 1 $\mu$bar & 1.74 \\
	0.1 $\mu$m 1000 cm$^{-3}$ tholin & CH$_{4}$ & 1 $\mu$bar & 1.48 \\
	1.0 $\mu$m 0.01 cm$^{-3}$ tholin & CH$_{4}$ & 1 $\mu$bar & 1.50 \\
	1.0 $\mu$m 0.1 cm$^{-3}$ tholin & CH$_{4}$ & 1 $\mu$bar & 1.51 \\
	\hline
	\hline
	\end{tabular}
	\label{Table:AtmosphereModel}
\end{table*}

\section{Conclusions}

In this work, we observed and studied a transiting hot Neptune, GJ3470b, which is the first sub-Jovian planet with detected Rayleigh scattering. Optical multi-filter observations of the exoplanet were obtained with the 2.4-m and 0.5-m telescopes at the Thai National Observatory (TNO) and the 0.6-m telescope at Cerro Tololo Inter-American Observatory (CTIO) in 2013-2016. Ten transit light curves were obtained and analyzed using the \texttt{TAP} program \citep{ago2005}. The derived data from the analysis provide a planet mass $M_{p} = 13.9\pm1.5 M_\oplus$, radius $R_{p} = 4.57\pm0.18 R_\oplus$, period $P = 3.3366496^{\mathrm {+0.0000039}}_{\mathrm {-0.0000033}}$ d, and inclination $i = 89.13^{\mathrm {+0.26}}_{\mathrm {-0.34}}$ degrees. A new ephemeris for GJ3470b is also provided. 

We perform the TTV analysis with the \texttt{TTVFaster} code of \cite{ago2016}, in order to determine an upper mass limit for a second planet in the system. The TTV signal indicates little variation, which excludes the presence of a hot Jupiter with orbital period less than 10 d in the system. The mutual Hill sphere also excludes the presence of a nearby planet with orbital period between 2.5 and 4.0 d.

For the transmission spectroscopy analysis, GJ3470b's low atmosphere mean molecular weight ($\mu = 1.08\pm0.20$) is obtained from the Rayleigh scattering fitting of the planet-star radius ratio variation in the optical. We confirm the steep Rayleigh scattering slope favoured by previous studies. Previous near-infrared data favour a methane atmosphere with high particle abundance (100 cm$^{-3}$ of tholin or polyaceylen) at high altitude (1000-1 $\mu$bar) when compared to the model atmosphere of \cite{how2012}. However, the models do not fit the data at optical wavelengths, which might be a consequence of the single atmosphere composition within the models. A mixed-ratio composition model could provide a better understanding of the planet's atmosphere.

\section*{Acknowledgements}

This work has made use of data obtained at the Thai National Observatory on Doi Inthanon, operated by NARIT, and the PROMPT-8 telescope, operated by the Skynet Robotic Telescope Network. SA gratefully acknowledges the support from the Thai Government Scholarship and the University of Manchester President's Doctoral Scholar Award. VSD acknowledges the support of the Science and Technology Facilities Council (STFC) and the Leverhulme Trust for ULTRASPEC operations. TRM acknowledges the support of STFC through grant ST/L0007333.




\bibliographystyle{mnras}
\bibliography{GJ3470refs}







\bsp	
\label{lastpage}
\end{document}